\documentclass{emulateapj}

\usepackage{apjfonts}

\slugcomment{submitted to the Astrophysical Journal, Letters}

\shorttitle{Supersoft X-ray Phase of V5583 Sgr}
\shortauthors{Hachisu et al.}

\begin{document}


\title{A Prediction of Supersoft X-Ray Phase of Classical Nova
V5583 Sagittarii}


\author{Izumi Hachisu}
\affil{Department of Earth Science and Astronomy,
College of Arts and Sciences,
University of Tokyo, Komaba 3-8-1, Meguro-ku, Tokyo 153-8902, Japan}
\email{hachisu@ea.c.u-tokyo.ac.jp}

\author{Mariko Kato}
\affil{Department of Astronomy, Keio University, 
Hiyoshi 4-1-1, Kouhoku-ku, Yokohama 223-8521, Japan}
\email{mariko@educ.cc.keio.ac.jp}

\author{Seiichiro Kiyota}
\affil{VSOLJ, Matsushiro 4-405-1003, Tsukuba 305-0035, Japan}
\email{skiyota@nias.affrc.go.jp}

%

\and

\author{Hiroyuki Maehara}
\affil{Kwasan Observatory, Graduate School of Science, Kyoto University,
Ohmine-cho Kita Kazan 17, Yamashina-ku, Kyoto 607-8471, Japan}
\email{maehara@kwasan.kyoto-u.ac.jp}

%



\begin{abstract}
We have observed the fast nova V5583 Sagittarii 
with five $B$, $V$, $y$, $R_C$, and $I_C$ bands,
and found that these multi-band light curves are almost identical
with those of V382 Vel 1999 until at least $\sim 100$ days after outburst.
A supersoft X-ray phase of V382 Vel was detected with {\it BeppoSAX} 
about six months after outburst.  V5583 Sgr outbursted a few days ago
the discovery on 2009 August 6.5 UT near its optical peak.  From a
complete resemblance between these two nova light curves,
we expect a supersoft X-ray phase of V5583 Sgr six months
after outburst.  Detection of supersoft X-ray 
turn-on/turnoff dates strongly constrain the evolution of a nova and,
as a result, mass range of the WD.
For a timely observation of a supersoft X-ray phase of V5583 Sgr,
we have calculated nova outburst evolution based on
the optically thick wind theory, which predicts the supersoft
X-ray phase: it will most probably start between days 100 and 140
and continue until days $200-240$ after outburst.
We strongly recommend multiple observations during 2009 December,
and 2010 January, February, and March to detect the turn-on and turnoff
times of the supersoft X-ray phase of V5583 Sgr.
\end{abstract}


\keywords{binaries: close --- novae, cataclysmic variables ---
stars: individual (V382 Velorum, V5583 Sagittarii) --- white dwarfs
--- X-rays: stars}



\section{Introduction}
     Classical novae are a thermonuclear runaway event on
a white dwarf (WD) in a binary system, in which the WD
accretes hydrogen-rich matter from the companion star.
When the accreted matter reaches a critical value, hydrogen
at the bottom of the WD envelope ignites to trigger a shell flash.
Just after the nova outburst, the envelope on the WD
rapidly expands to a giant size and optically thick winds blow.
Then the photosphere gradually shrinks whereas the total luminosity
is almost constant during the outburst.  Thus, the photospheric
temperature $T_{\rm ph}$ increases with time.  The main emitting
wavelength region moves from optical to ultraviolet and
finally to supersoft X-ray \citep[e.g.,][]{kat94h}.

Thus, classical novae become a transient supersoft
X-ray source in a later phase of the outburst,
but their X-ray detections are rather rare mainly
because of sparse observing time of X-ray satellites
\citep[e.g.,][]{kra96, ori01a, nes07a}.  If the turn-on/turnoff dates
of the supersoft X-ray are detected, we are able to constrain 
the evolution of hydrogen shell-burning on the WD and, as a result,
the mass range of the WD \citep[e.g.,][]{hac06kb, hac09k, hac10ka,
 hac07kl}.  Therefore detection of a supersoft X-ray phase of a nova
provides us with rich information on the WD.

V5583 Sgr is a fast classical nova, discovered on 2009 August 6.5 UT
by K. Nishiyama and F. Kabashima at mag $\sim 7.7$ \citep{nis09}.
Their survey frames on July 22.5 and 29.6 UT showed nothing
at this position (limiting mag 12.7), and nothing is visible on
Digitized Sky Survey images.  The ASAS3V images showed $V = 7.78$
on August 6.2 UT \citep{nis09}.

We started multi-band photometric observation of V5583 Sgr from
one day after the discovery, i.e., from August 7.5 UT.  To our surprise,
the observed multi-band light curves are almost identical
with those of V382 Vel, which is also a fast classical nova outbursted
in 1999.  A supersoft X-ray phase of V382 Vel was clearly detected
with the X-ray satellite {\it BeppoSAX} about six months after
the outburst.  From the perfect resemblance between these two novae,
we expect a supersoft X-ray phase of V5583 Sgr similar to that of
V382 Vel.
In this Letter, we have calculated a supersoft X-ray phase
for V5583 Sgr, and predict turn-on/turnoff dates for timely detection.

     In the next section (Section \ref{observation}),
we briefly describe our multi-band photometric observation of V5583 Sgr.
Section \ref{opticall_thick_wind_model} introduces our model of 
nova light curves based on the optically thick wind theory and summarizes
numerical results for prediction of a supersoft X-ray phase.
Discussion follows in Section \ref{discussion}.

\section{Observation} \label{observation}
Optical observation was started one day after the discovery 
\citep{nis09}.  Each observer and their observational
details are listed in Table \ref{observers}.  
Maehara started observation on August 7 and obtained 23 nights data
for five bands of $B$, $V$, $y$, $R_C$, and $I_C$ (until 2009
November 3).  Kiyota obtained four bands of $B$, $V$, $R_C$, and $I_C$
for 14 nights starting from 2009 August 8 (until 2009 October 18).
The magnitudes of this object were measured by using the local
standard star, HD 321237 with $V = 11.683$ and $B-V = +0.198$
(Kiyota) from AAVSO (the American Association of Variable
Star Observers), or TYC 7395-2150-1 with $V=9.80$ and
$B-V=+0.50$ (Maehara) from Tycho catalog.


\begin{deluxetable}{llllr}
\tabletypesize{\scriptsize}
\tablecaption{Observations \label{observers}}
\tablewidth{0pt}
\tablehead{
\colhead{name of} & \colhead{location} & \colhead{telescope} & 
\colhead{observed} & \colhead{No. of obs.} \cr
\colhead{observer} & \colhead{} & \colhead{aperture} & 
\colhead{bands} & \colhead{nights} 
}
\startdata
Kiyota & Mayhill, USA & 30cm & $B,V,R_c,I_c$ & 8 \cr
Kiyota & Moorook, Australia & 25cm & $B,V,R_c,I_c$ & 6 \cr
Maehara & Kyoto, Japan & 25cm & $B,V,y,R_c,I_c$ & 23
\enddata
\end{deluxetable}


\begin{figure*}
\epsscale{0.75}
\plotone{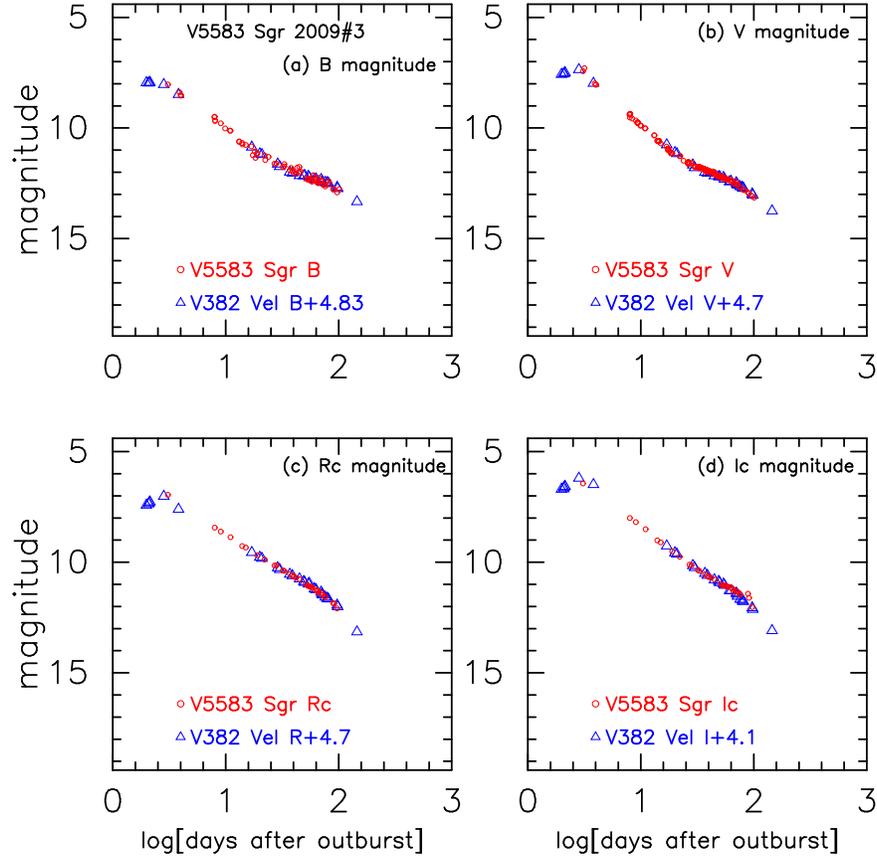}
\caption{
Our four multi-band ($B$, $V$, $R_C$, and $I_C$)
optical light curves for V5583 Sgr together with
those of V382 Vel 1999.  Observational V382 Vel data of $B$,
$V$, $R$, and $I$ are taken from IAU Circulars 7176, 7179, 7196, 7209,
7216, 7226, 7232, 7238, and 7277.  Here we assume the outburst day of 
$t_{\rm OB}=$JD 2455048.0 (2009 August 4.5 UT) for V5583 Sgr,
and $t_{\rm OB}=$JD 2451319.0 (1999 May 20.5 UT) for V382 Vel
\citep{hac10ka}.
\label{light_curve_v5583_sgr_v382_vel_4fig}
}
\end{figure*}

Our observational results are plotted in Figure
\ref{light_curve_v5583_sgr_v382_vel_4fig} 
for four bands of $B$, $V$, $R_C$, and $I_C$ of V5583 Sgr
together with the $B$, $V$, $R$, and $I$ light curves of V382 Vel.
We have added other observational points available in VSOLJ
(the Variable Star Observing League of Japan) and AAVSO archives.
It is very clear that these two nova light curves are almost 
identical with each other.  Here we assume that the outburst day
of V5583 Sgr is $t_{\rm OB}=$JD 2455048.0 (2009 August 4.5 UT).

From the almost complete resemblance between these two novae, we can
deduce various features of the classical nova V5583 Sgr.
(1) V5583 Sgr is probably a neon nova because V382 Vel was
identified as a neon nova \citep{woo99}.
(2) The chemical composition of nova ejecta is similar to that of
V382 Vel obtained by \citet{sho03} and \citet{aug03}.
(3) A supersoft X-ray phase will be detected about
six months after the outburst similarly to the V382 Vel case
\citep{ori02}. (4) The interstellar
extinction is calculated to be $E(B-V)=0.33$ from
\begin{eqnarray}
& & \left [ E(B-V) \right ]_{\rm V5583~Sgr} 
- \left [ E(B-V) \right ]_{\rm V382~Vel} \cr
&=& \left [ (B-V) - (B-V)_0 \right ]_{\rm V5583~Sgr} 
- \left [ (B-V) - (B-V)_0 \right ]_{\rm V382~Vel} \cr
&=& \left( B \right)_{\rm V5583~Sgr} - \left( B \right)_{\rm V382~Vel} 
+ \left( V \right)_{\rm V5583~Sgr} - \left( V \right)_{\rm V382~Vel} \cr
&=& 4.83 - 4.7 = 0.13
\end{eqnarray}
together with $E(B-V)= 0.2$ for V382 Vel \citep{sho03}.  Here we
assume that the intrinsic color of $(B-V)_0$ is the same between
V5583 Sgr and V382 Vel.  (5) The distance modulus of V5583 Sgr is
estimated from the comparison between the two nova brightnesses.
Since the distance modulus of V382 Vel is already known to be
$(m-M)_V = 11.5 \pm 0.1$ \citep{hac10ka} and the difference between
the two nova brightnesses is $\Delta V = 4.7$ from 
Figure \ref{light_curve_v5583_sgr_v382_vel_4fig}b, we obtain
$(m-M)_V = 16.2 \pm 0.1$ for V5583 Sgr.
(6) Therefore the distance to V5583 Sgr is estimated to be
$d \sim 11 \pm 1$ kpc from $(m-M)_V = 5 \log (d/10) + A_V$ and 
$A_V = 3.1 E(B-V) = 1.0$.
These values are summarized in Table \ref{properties}.


\begin{deluxetable}{llll}
\tabletypesize{\scriptsize}
\tablecaption{Physical Properties of V5583 Sgr \label{properties}}
\tablewidth{0pt}
\tablehead{
\colhead{subject} & \colhead{symbol} & \colhead{} & \colhead{present work} 
}
\startdata
nova type & ... & ... & neon nova \cr
extinction & $E(B-V)$  & ... & 0.33 \cr
absorption in $V$-band & $A_V$  & ... & 1.0 \cr
distance modulus & $(m-M)_V$  & ... & $16.2 \pm 0.1$ \cr
distance & $d$  & ... & $11 \pm 1$ kpc \cr
WD mass & $M_{\rm WD}$  & ... & $1.23 \pm 0.05~M_\sun$ \cr
supersoft X-ray on & $t_{\rm X-on}$  & ... & $120 \pm 20$ days \cr
supersoft X-ray off & $t_{\rm X-off}$  & ... & $220 \pm 20$ days
\enddata
\end{deluxetable}





\section{Model Light Curves and Supersoft X-ray Phase}
\label{opticall_thick_wind_model}

The decay phase of novae can be
followed by a sequence of steady-state solutions \citep[e.g.,][]{kat94h}.
Using the same method and numerical techniques as in \citet{kat94h},
we have followed evolutions of novae by connecting steady state solutions
along the decreasing envelope-mass sequence.  The mass of the
hydrogen-rich envelope is decreasing due to wind mass-loss
and nuclear burning.  We solve a set of equations, consisting of
the continuity, equation of motion, radiative diffusion,
and conservation of energy, from the bottom of the hydrogen-rich
envelope through the photosphere assuming spherical symmetry.
Winds are accelerated deep inside the photosphere so that
they are called ``optically thick winds.''

We have calculated nova light curves of V5583 Sgr 
in the same way as for V382 Vel \citep{hac10ka}.
Supersoft X-ray light curves are calculated assuming blackbody 
spectrum with the photospheric temperature, $T_{\rm ph}$, for
the energy range of 0.2--0.6 keV \citep[see, e.g.,][]{hac09k}.
The UV 1455\AA\  band is also useful to follow nova evolutions
and to determine WD masses
\citep[e.g.,][]{cas02, hac06kb, hac10ka, kat09hc},
although they are not available both for V382 Vel and V5583 Sgr.

For optical and near IR light curves, flux at the frequency
$\nu$ is estimated from free-free emission spectrum,
Equation (9) of \citet{hac06kb}, i.e.,
\begin{equation}
F_\nu \propto {{\dot M_{\rm wind}^2} \over {{v_{\rm ph}^2 R_{\rm ph}}}},
\label{wind-free-free-emission}
\end{equation}
during the optically thick wind phase,
where $\dot M_{\rm wind}$ is the wind mass-loss rate,
$v_{\rm ph}$ the wind velocity at the photosphere, and
$R_{\rm ph}$ the photospheric radius, all of which are taken
from our optically thick wind solutions.
After the optically thick wind stops, the total mass of the
ejecta remains constant in time.  The flux from such homologously
expanding ejecta is estimated from Equation (19) of \citet{hac06kb},
i.e.,
\begin{equation}
F_\nu \propto t^{-3},
\label{expansion-free-free-emission}
\end{equation}
where $t$ is the time after the outburst.

\begin{figure}
\epsscale{1.15}
\plotone{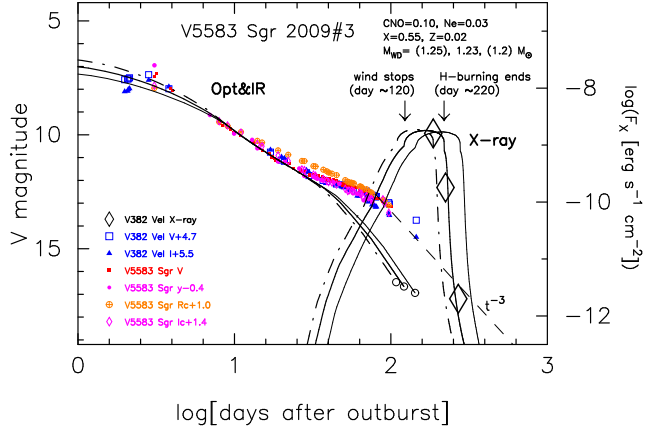}
\caption{
Optical and supersoft X-ray light curves for V5583 Sgr together with
V382 Vel.  We plot free-free emission model light curves (labeled
``Opt\&IR'') and $0.2-0.6$~keV supersoft X-ray fluxes (labeled
``X-ray'') of three WD mass models of $M_{\rm WD}= 1.25 ~M_\sun$
 ({\it thick dash-dotted line}),
$1.23 ~M_\sun$ ({\it thick solid line}), and $1.2 ~M_\sun$
({\it thin solid line}),
for the envelope chemical composition of $X=0.55$, $Y=0.30$,
$Z=0.02$, $X_{\rm CNO}= 0.10$, $X_{\rm Ne}= 0.03$.
We select the $1.23 ~M_\sun$ WD as a best model reproducing
the supersoft X-ray data of V382 Vel.
The X-ray absorbed flux data ({\it large open diamonds})
of V382 Vel are taken from \citet{ori02} and \citet{bur02}.
Large open circles at the right edge of each free-free light curve
correspond to the end epoch of an optically thick wind phase,
during which the free-free flux is calculated by
Equation (\ref{wind-free-free-emission}). 
\label{all_mass_v5583_sgr_v1500_cyg_x55z02o10ne03}
}
\end{figure}

We assume that the chemical composition of hydrogen-rich envelope
is similar to that of V382 Vel and adopt
a set of $X=0.55$, $Y=0.30$, $Z=0.02$, $X_{\rm CNO}= 0.10$, 
$X_{\rm Ne}= 0.03$, based on the composition analyses for V382 Vel
by \citet{sho03} and by \citet{aug03}
\citep[see Table 1 of][]{hac06kb}.  
We plot our model free-free and X-ray light curves
in Figure \ref{all_mass_v5583_sgr_v1500_cyg_x55z02o10ne03}.
\citet{hac10ka} estimated the WD mass of V382 Vel to be
$M_{\rm WD}= 1.23 \pm 0.05 ~M_\sun$ 
and the supersoft X-ray turn-on/off times of 
$t_{\rm X-on} \sim 120$ days and $t_{\rm X-off} \sim 220$ days
as shown in Figure \ref{all_mass_v5583_sgr_v1500_cyg_x55z02o10ne03}
and as listed in Table \ref{properties}.  Errors come from ambiguity
of the chemical composition. 

Because the two nova light curves are almost identical, our light curve
fitting of V5583 Sgr gives similar results.
We obtain the supersoft X-ray phase of V5583 Sgr to be days
120--220, i.e., the supersoft X-ray phase lasts from early December
of 2009 until mid March of 2010.
Since the supersoft X-ray phase has already started in our
estimate, urgent X-ray observation is required.


\section{Discussion}
\label{discussion}



\begin{deluxetable*}{llclllr}
\tabletypesize{\scriptsize}
\tablecaption{Distance and Absorption of Novae\tablenotemark{a}
\label{distance_moduli_of_novae}}
\tablewidth{0pt}
\tablehead{
\colhead{object} &
\colhead{...} &
\colhead{$(m-M)_V$\tablenotemark{b}} &
\colhead{$A_V$} &
\colhead{distance} &
\colhead{discovery} &
\colhead{ref.\tablenotemark{c}} \\
\colhead{} &
\colhead{} &
\colhead{} &
\colhead{} &
\colhead{(kpc)} &
\colhead{satellite} &
\colhead{}
}
\startdata
V1281 Sco 2007\#2 & ... & 17.8 & 2.17 & 13 & Swift & 1 \\
V458 Vul 2007 & ... & 17.0 & 1.86 & 11 & Swift & 2 \\
V597 Pup 2007\#1 & ... & 16.9 & 0.93 & 16 & Swift & 3 \\
V2467 Cyg 2007 & ... & 16.3 & 4.65 & 2.2 & Swift & 4 \\
V5116 Sgr 2005\#2 & ... & 16.2 & 0.81 & 12 & XMM Newton & 5 \\
V5583 Sgr 2009\#3 & ... & 16.2 & 1.02 & 11 & -- & 6 \\
V574 Pup 2004 & ... & 15.4 & 2.2 & 4.6 & Swift & 5 \\
V4743 Sgr 2002\#3 & ... & 13.8 & 0.78 & 3.8 & Chandra & 7 \\
V1494 Aql 1999\#2 & ... & 13.4 & 1.83 & 2.2 & Chandra & 8 \\
V1974 Cyg 1992 & ... & 12.3 & 1.00 & 1.8 & ROSAT & 9 \\
V598 Pup 2007\#2 & ... & 11.7 & 0.27 & 2.1 & XMM Newton & 10 \\
V382 Vel 1999 & ... & 11.4 & 0.62 & 1.5 & BeppoSAX & 11 
\enddata
\tablenotetext{a}{supersoft X-ray on/off detected novae except 
V5583 Sgr}
\tablenotetext{b}{distance modulus taken from Table 8 of \citet{hac10ka}}
\tablenotetext{c}{reference for $A_V$ or $E(B-V)$,
where we assume that $A_V = 3.1 E(B-V)$:
1-\citet{rus07a}, 
2-\citet{wes08}, 
3-\citet{nes08c}, 
4-\citet{maz07}, 
5-\citet{bur08}, 
6-present work, 
7-\citet{van07}, 
8-\citet{iij03}, 
9-\citet{cho93}, 
10-\citet{rea08}, 
11-\citet{sho03}, 
}
\end{deluxetable*}

The distance modulus of V5583 Sgr can also be estimated from
the comparison with the absolute magnitude of free-free emission
model light curves.  \citet{hac10ka} obtained the absolute magnitudes
at the points denoted by open circles of model light curves, which 
correspond to the end of an optically thick wind phase.  For the model
of $1.23 ~M_\sun$ WD, its absolute magnitude is $M_{\rm w}= 0.5$
\citep{hac10ka}.  This point corresponds to $m_{\rm w}= 16.7$
in Figure \ref{all_mass_v5583_sgr_v1500_cyg_x55z02o10ne03}.  
Thus we have
\begin{equation}
(m-M)_V = m_{\rm w} - M_{\rm w} = 16.7 - 0.5 = 16.2.
\end{equation}
This value is consistent with $(m-M)_V=16.2 \pm 0.1$ estimated with the
difference between V382 Vel and V5583 Sgr in the previous section.

Supersoft X-rays are heavily absorbed by interstellar (or circumstellar)
neutral hydrogen.  Therefore the detection of supersoft X-rays depends 
not only on the distance ($d$) but also on the absorption ($A_V$).
Table \ref{distance_moduli_of_novae} lists 11 novae with a
supersoft X-ray phase being detected in the order of
decreasing distance modulus.  The position of V5583 Sgr is mid of
the list, that is, neither the distance nor the absorption is 
too large to be detected.
We expect detection of supersoft X-rays from V5583 Sgr.



\acknowledgments

We thank the Variable Star Observing League of Japan (VSOLJ)
and the American Association of Variable Star Observers (AAVSO)
for the optical photometric data on V5583 Sgr.
This research has been supported in part by Grants-in-Aid for
Scientific Research (20540227)
of the Japan Society for the Promotion of Science.

\end{document}